\documentstyle[12pt,twoside,fleqn,espcrc1,epsf,epsfig,psfig]{article}
\newcommand{\beq}{\begin{equation}}
\newcommand{\eeq}{\end{equation}}
\newcommand{\beqa}{\begin{eqnarray}}
\newcommand{\eeqa}{\end{eqnarray}}

\newcommand{\AmS}{{\protect\the\textfont2
  A\kern-.1667em\lower.5ex\hbox{M}\kern-.125emS}}

\title{Baryon form factors: Model--independent results\footnote{
Plenary talk presented at {\it "Nucleon '99"}, Frascati, Italy, June
7-9, 1999.}
}  
\author{Ulf-G. Mei{\ss}ner\address{Forschungszentrum J\"ulich, 
Institut f\"ur Kernphysik (Theorie), 
D-52425 J\"ulich, Germany}}
\begin{document}

\maketitle

\vspace{-5.2cm} {\flushright {\small FZJ-IKP(TH)-1999-18}\\[0.2em]

{\hfill  {\small {\tt hep-ph/9907323}}}}%

\vspace{5cm} 

\begin{abstract}
\noindent Baryon form factors can be analyzed in a largely model--independent
fashion in terms of two complementary approaches. These are  chiral perturbation
theory and dispersion relations. I review the status of dispersive calculations
of the nucleon electromagnetic form factors in the light of new data. Then, I present
the leading one--loop chiral perturbation theory analysis of the hyperon and the
strange nucleon form factors. Open problems and challenges are also discussed. 
\end{abstract}

\section{Outline}

\noindent There are many interesting recent theoretical  developments
concerning the strange and electromagnetic (em) form factors (ffs) of the nucleon
and of the hyperons. Although often useful, in this talk I will eschew models and only
discuss some largely model--independent results which have emerged over the
last years. The pertinent methods are {\it dispersion} {\it relations} and {\it
chiral} {\it perturbation} {\it theory}. For such a model--independent description,
one  has of course to pay a price, namely one needs a certain amount of input data.
Indeed, dispersion relations make use of all available data and can be applied
over the full range of accessible energies. Chiral effective field theories are limited
to energies below the typical hadronic scale of about 1~GeV. As I will show, both
methods allow to deepen our understanding of hadronic structure.  
Instead of a general introduction, I will briefly summarize which 
topics will be addressed in this talk.
{\it (i)~Dispersion theory for the nucleon em ffs}: Dispersion
relations have been used for many decades to analyze the data on electron--proton
(or deuteron) scattering. The status of such calculations as of 1996 is
reviewed in the talk~\cite{ulfdaphce}. Here, I will briefly discuss the
impact of the recent data from MAMI, NIKHEF, BATES and CEBAF on these calculations.  
{\it (ii)~Hyperon form factors}: Recent measurements of the
$\Sigma^-$ radius using elastic hadron--electron scattering at CERN and
Fermilab have triggered a chiral perturbation theory (CHPT) analysis of the 
hyperon form factors. It seems to indicate that for these observables
SU(3) baryon CHPT is indeed a very {\it effective} method.
{\it (iii)~Strangeness in the nucleon}: Recent data from MIT-BATES
and Jefferson Lab allow for a complete leading one--loop analysis of the strange
nucleon form factors in the framework of baryon CHPT. This has been attempted
before but could not be done due to the lack of data.
In the following, I will address these questions and outline further
directions of theoretical research.

\section{Dispersion relations update}

\noindent The structure of the nucleon as probed with virtual photons
is parametrized in terms of four form factors. One can either
choose the basis of the Dirac and Pauli ffs ($F_{1,2} (t)$), or equivalently,
the so--called Sachs (electric and magnetic) ffs ($G_{E,M}(t)$), with 
$t=q^2=-Q^2$ the invariant momentum transfer squared (note that $t<0$ in
electron scattering). In the Breit frame,
the Sachs form factors give the distribution of charge and magnetization
within the proton and the neutron. The neutron electric ff plays a 
particular role since the neutron charge is zero, but still there is a
non--vanishing distribution of charge which leads to the non--vanishing
but small ff. Although 
not proven strictly (but shown to hold in all orders in perturbation
theory), one can write down an unsubtracted dispersion relation for $F(t)$ (which
is a generic symbol for any one of the four ff's),
\beq
F(t) = \frac{1}{\pi} \int_{t_0}^\infty \, dt' \, \frac{{\rm Im} \, F(t)}{t'-t}
\, \, , \eeq
with $t_0$ the two (three) pion threshold for the isovector (isoscalar) ffs.
Im~$F(t)$ is called the {\it spectral}
{\it  function}. It is advantageous to work in
the isospin basis, $F_i^{I=0,1} = (F_i^p \pm F_i^n)/2$, since the photon
has an isoscalar ($I=0$) and an isovector ($I=1$) component. 
In general, the spectral functions 
can be thought  of as a superposition of vector meson poles and some 
continua, related to n-particle thresholds, like
e.g. $2\pi$, $3\pi$, $K \bar{K}$, $N\bar{N}$ and so on. 
For example, in the Vector Meson
Dominance (VMD) picture, one simply retains a set of poles. This turns out
to be an insufficient approximation.
The dispersive approach of refs.\cite{mmd,hmd} includes three (or four) isoscalar 
and three isovector poles. Unitarity allows to reconstruct the isovector
spectral functions up to $t \simeq 1\,$GeV$^2$ and this model--independent
piece contains the $\rho$. In addition, perturbative QCD behaviour plus some 
other refinements can be built in. 
The proton electric and magnetic form factors
in the space-- and time--like as well as in the unphysical region
($0\le t \le 4m^2$, with $m$ the nucleon mass)
for this fit are shown in figs.1,2. 
\begin{figure}[b]
\begin{minipage}[h]{77mm}
\psfig{figure=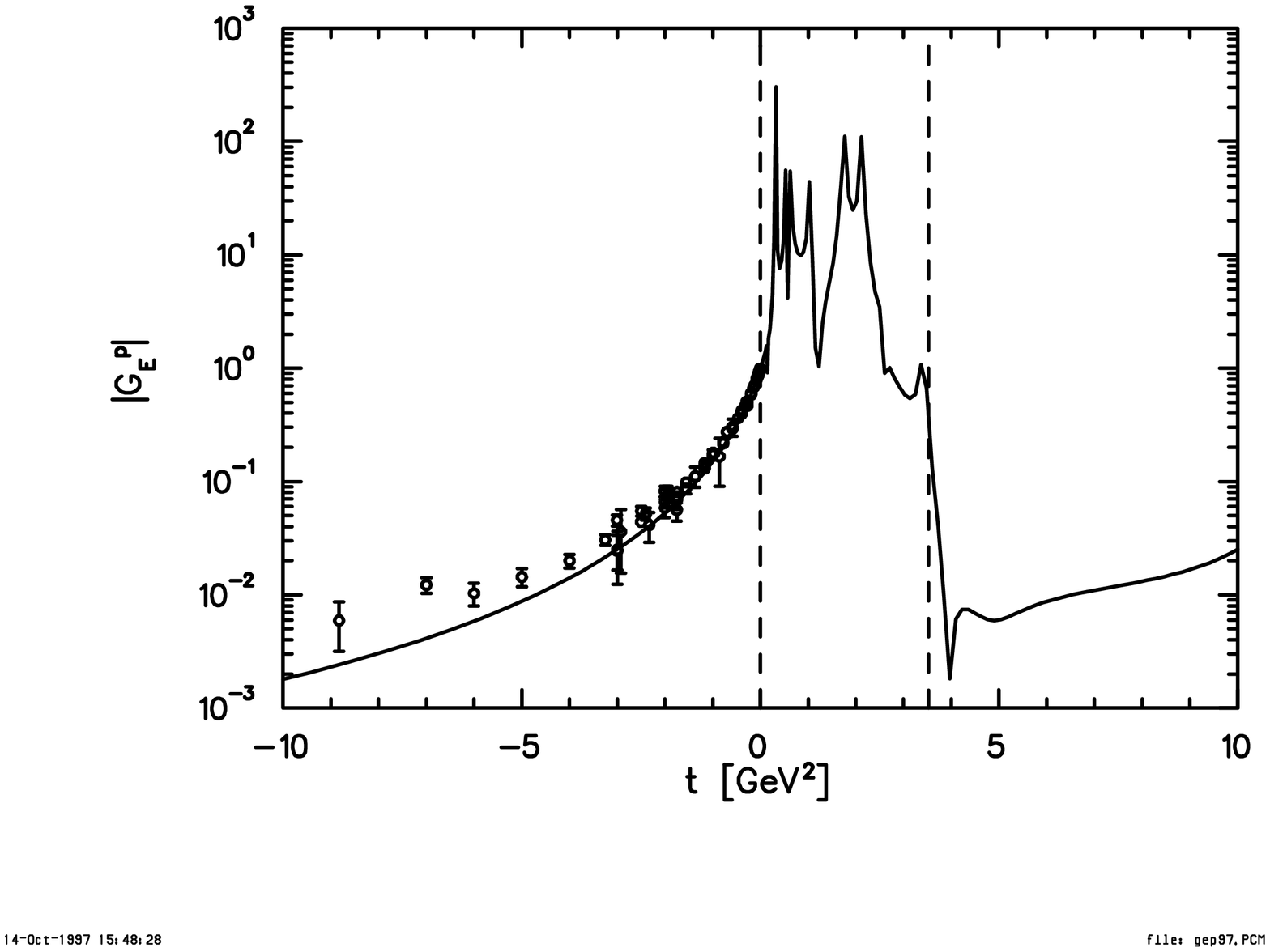,height=7cm,angle=90}
\vspace{-0.9cm}
\caption{Dispersive analysis of the electric proton form factor (solid
  line).}
\label{fig:gep}
\end{minipage}
\hspace{\fill}
\begin{minipage}[h]{77mm}
\psfig{figure=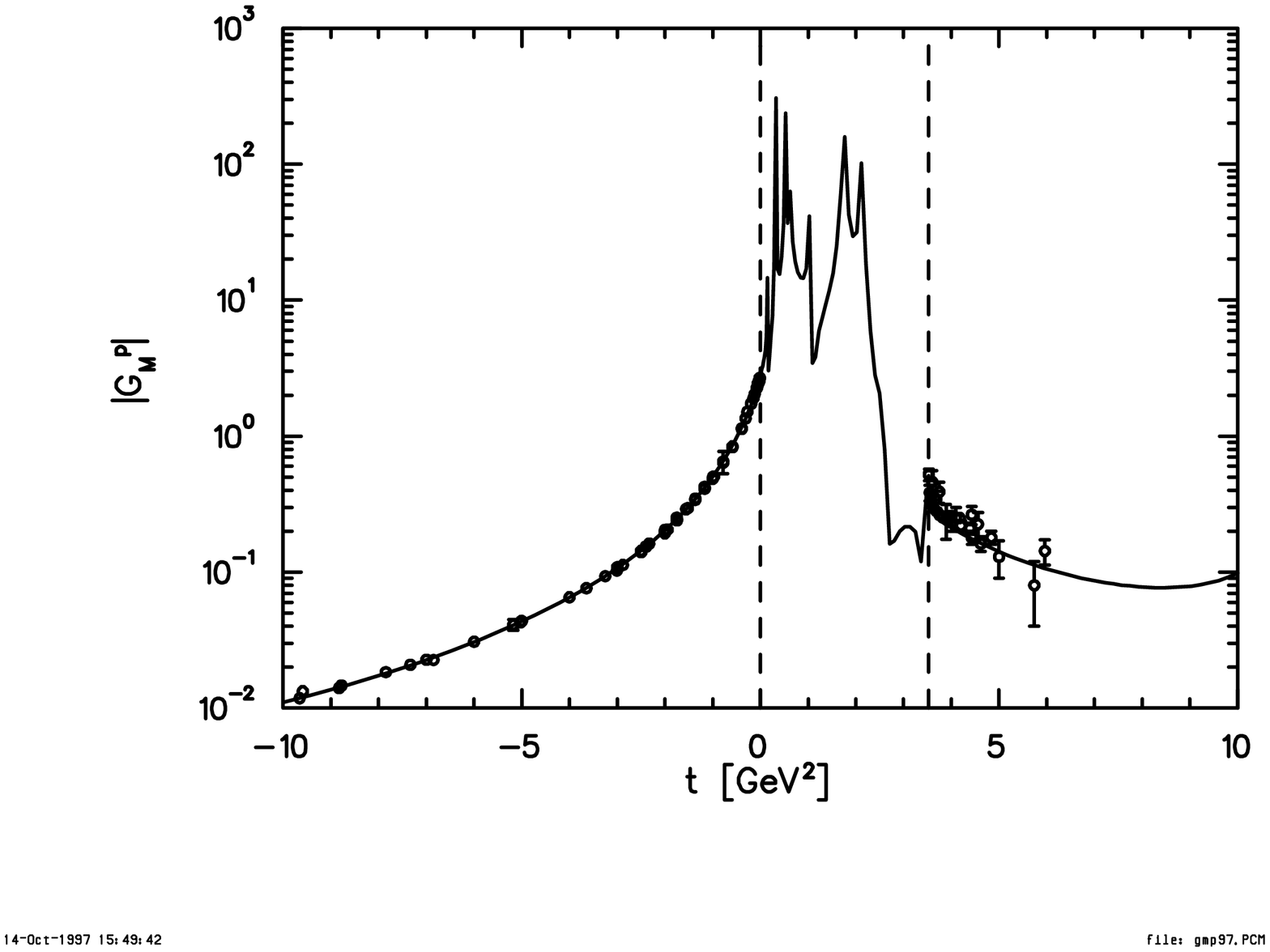,height=7cm,angle=90}
\vspace{-0.9cm}
\caption{Dispersive analysis of the magnetic proton form factor (solid
  line).}
\label{fig:gmp}
\end{minipage}
\end{figure}
\noindent
It should be
pointed out that for the existing data scaling as predicted by pQCD
is not yet observed, in particular for the ratio $Q^2 F_2^p (Q^2)/F_1^p (Q^2)$
below $Q^2 = 10\,$GeV$^2$. In that region, there is still a sizeable vector meson
pole contribution.  
Recently, a different dispersive approach has been presented
which allows for a consistent description of the time-- and space--like data
but  less theoretical constraints are built in~\cite{bald}. Most interesting
is the appearance of a resonance--like structure just below the two--nucleon
threshold in the description of the neutron data. An interpretation in terms
of a dibaryon could be possible, but clearly this phenomenon deserves
more study. It would also be interesting
to see a refined version of this calculation.
\begin{figure}[ht]
\begin{minipage}[htb]{77mm}
\vspace{.5cm}
\psfig{figure=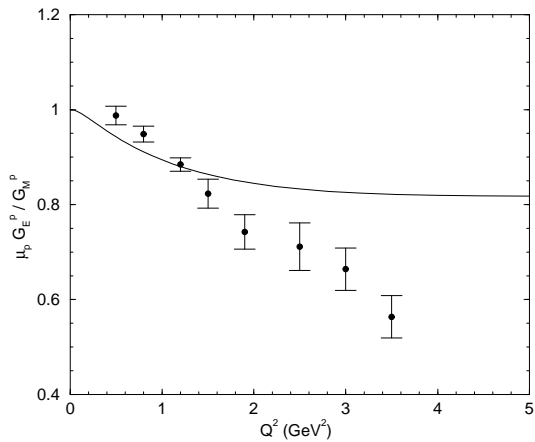,width=7.cm}
\caption{The ratio $\mu_p G_E^p (Q^2)/G_M^p (Q^2)$ as measured at Jefferson
Lab. The solid line is the result of a dispersive analysis of {\it all} data
on {\it all} nucleon form factors. For further discussions, see the text.}
\label{fig:jlab}
\end{minipage}
%
\hspace{\fill}
\begin{minipage}[ht]{77mm}
Since the time of ref.\cite{ulfdaphce}, 
new data for the ratio of the protons electric and magnetic 
form factors have been obtained at CEBAF 
using polarization transfer~\cite{jlabdat} (for $Q^2$ between 0.5 and 3.5~GeV$^2$.).
These data are much more precise than previous ones since
the Rosenbluth separation is avoided. They are shown in fig.3. The
solid line gives the result of the dispersive analysis with at most
three isovector and four isoscalar poles. Even if one decreases the
experimental errors to zero, the result of the fit does not change. I conclude
that in the {\it ansatz} for the spectral functions an essential physics ingredient
is missing. Whether this can be remedied by the inclusion of further poles or
a better treatment of continua (beyond the two--pion continuum) is an open problem
which requires more theoretical work. 
\end{minipage}
\end{figure}

\noindent The situation concerning the neutron electric form factor is not yet 
satisfactory. The extraction of this quantity from elastic electron scattering
off the deuteron was claimed to be plagued by a strong dependence on the
deuteron wavefunction~\cite{pla}, as shown in fig.4 by the thin dotted
lines. However, in view of the fact that all modern two--nucleon potentials
give the same results for a huge amount of two-- and three--nucleon observables,
this wavefunction dependence is presumably an artefact of the treatment in
ref.\cite{pla}. A modern extraction from the same data based on the Argonne
V18 potential is shown by the open circles in fig.4~\cite{rbw}. Also shown
are the recent data from BATES, MAMI and NIKHEF~\cite{newnff}. 
These measurements invoke polarized
targets or polarization transfer and thus have much less systematic uncertainties.
Still, the error bars are sizeable. Note that these new data tend to give larger
values for the neutron electric form factor, with the exception of the two
points at $Q^2=0.32$ and $0.36\,$GeV$^2$ obtained with the help of a polarized
$^3$He target. The solid line in that figure is the result
of the dispersive analysis. We took the Saclay data as given by the Paris potential
but doubled the error bars to account for the wavefunction dependence. A systematic
treatment of these data using all available high precision potentials and consistent
exchange currents is called for. I expect that such an investigation will lead to
a sizeably reduced  wavefunction
dependence (if there is any). Of course, a reduction of the error bars for the
polarization data would also be helpful to further pin down the
neutron electric form factor. In particular, there is still the discrepancy between 
the deuteron and the helium points to be resolved. 

\vspace{-.3cm}
\begin{figure}[ht]

\hskip 1.1in
\psfig{figure=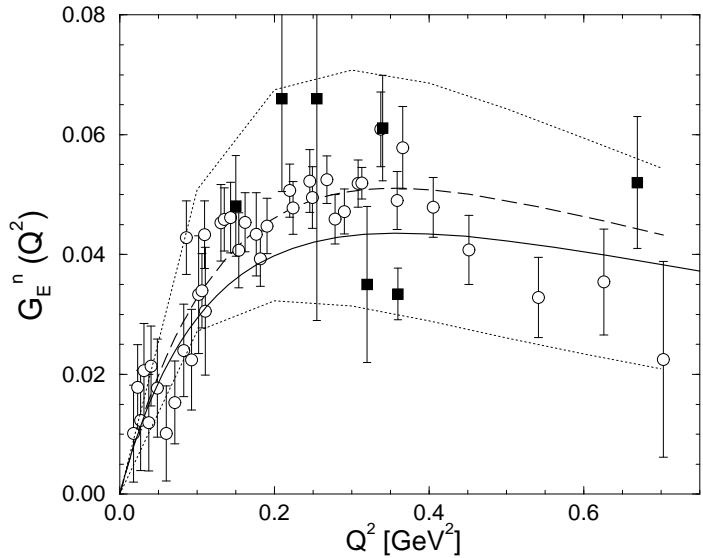,height=3in}
\vspace{-1.cm}
 \caption{The neutron electric form factor. For explanations, 
 see the text.\label{fig:Gen}}
\end{figure}

\section{Hyperon form factors}

\noindent
To third order in the chiral expansion, i.e. to leading one--loop order,
the electromagnetic form factors of the nucleon have been studied in
refs.\cite{gss,bkkm,bfhm}. At that order, one has to deal with two
counterterms in the electric  and two in the magnetic ffs. 
Using e.g. the proton and neutron electric radii and magnetic moments as input, the
ffs are fully determined to that order. In particular, no counterterms appear
in the momentum expansion of the magnetic ffs. To this order in the chiral
expansion, the ffs are precisely described for momentum transfer squared
up to $Q^2 \simeq 0.2\,$GeV$^2$. It appears therefore natural
to extend such an investigation to the three flavor case. Surprisingly,
that has never been attempted until recently~\cite{khm}
despite a huge amount of studies in three flavor chiral perturbation theory.
This investigation was triggered by the recent results on the $\Sigma^-$ radius
obtained by the WA89 collaboration at CERN and by the SELEX collaboration at
FNAL (note that the SELEX results are still preliminary),
\begin{equation}
\langle r^2_{\Sigma^-}\rangle_{\exp} 
= 0.92\pm 0.32\pm 0.40~{\rm fm}^2~\cite{WA89}~,\quad
\langle r^2_{\Sigma^-}\rangle_{\exp} 
= 0.60\pm 0.08\pm 0.08~{\rm fm}^2~\cite{SELEX}~,
\end{equation}
obtained by scattering a highly boosted hyperon beam in the electronic cloud of a heavy
atom (elastic hadron--electron scattering). The pattern of the charge radii
embodies information on SU(3) breaking and the structure of the groundstate octet.
In a CHPT calculation of the corresponding ffs, the baryon structure is to some 
part given by the meson (pion and kaon) cloud and in part by shorter distance physics 
parametrized in terms of local contact interactions. In the general case, 
such a splitting depends on the regulator scheme and scale one chooses. Here, we work in 
standard dimensional
regularization and set $\lambda =1\,$GeV throughout (since this is the natural
hadronic scale). If one performs the SU(3) calculation to third order, one
has no new counterterms as compared to the SU(2) calculation. Therefore, fixing
the low--energy constants (LECs) from proton and neutron properties allows one
to make parameter--free predictions for the hyperons. As an added bonus, kaon
loops induce a momentum dependence in the isoscalar magnetic form factor of the
nucleon, as first pointed out in ref.\cite{hms}, whereas in the pure SU(2)
calculation, $G_M^{I=0} (Q^2)$ is simply constant.  This allows one to study the
contribution of kaon loops (strangeness) to the em ffs of the nucleon 
(not to be confused with the strange ffs to be discussed below). 

\medskip
\begin{figure}[h]
\begin{minipage}[htb]{77mm}
\psfig{figure=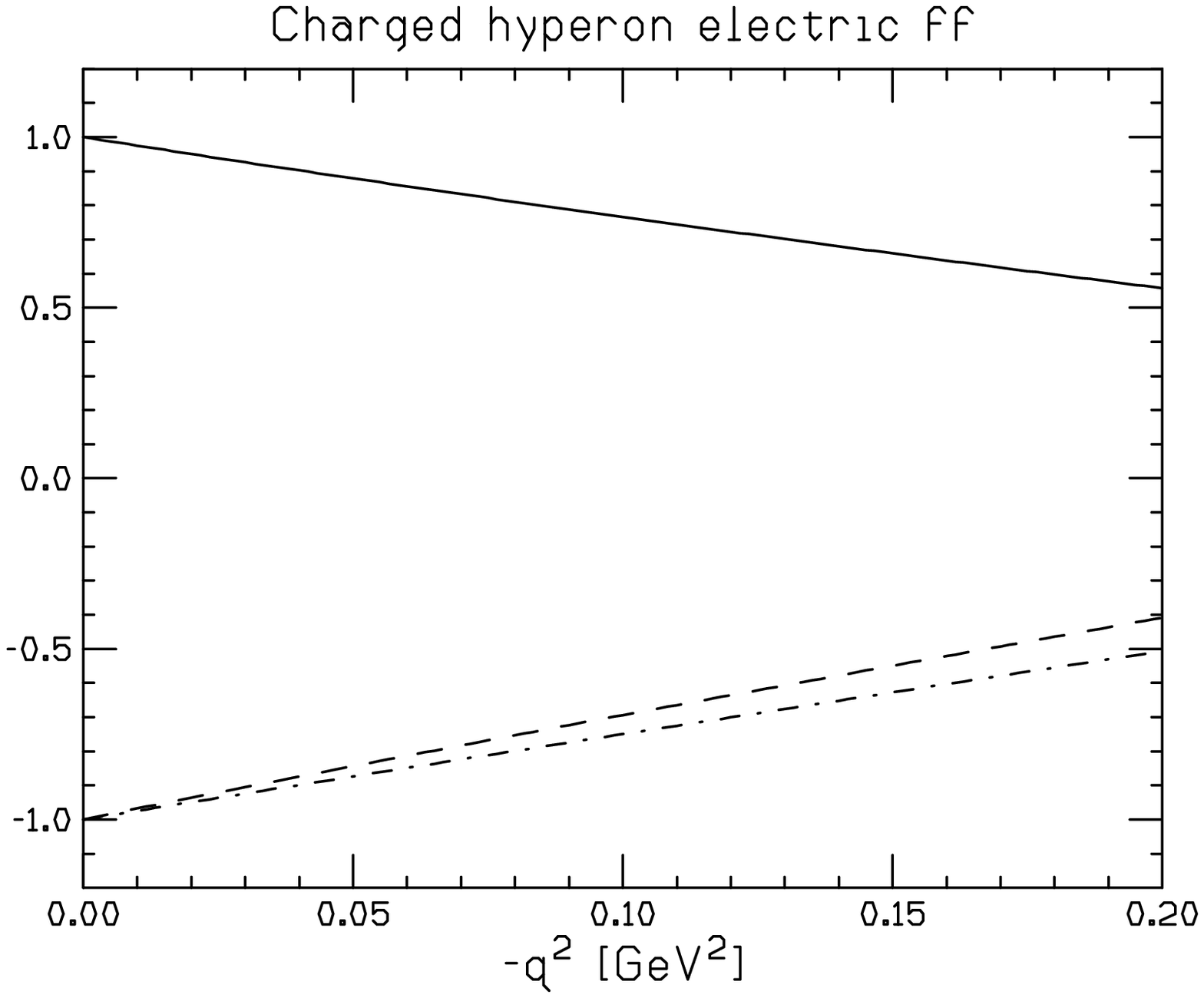,width=6.8cm}
\caption{The electric form factors of the charged hyperons calculated in
three flavor baryon CHPT.
Solid, dashed, dot--dashed line: $\Sigma^+$, $\Sigma^-$, $\Xi^-$, in order.}
\label{fig:hyelff}
\end{minipage}
\hspace{\fill}
\begin{minipage}[ht]{77mm}
\psfig{figure=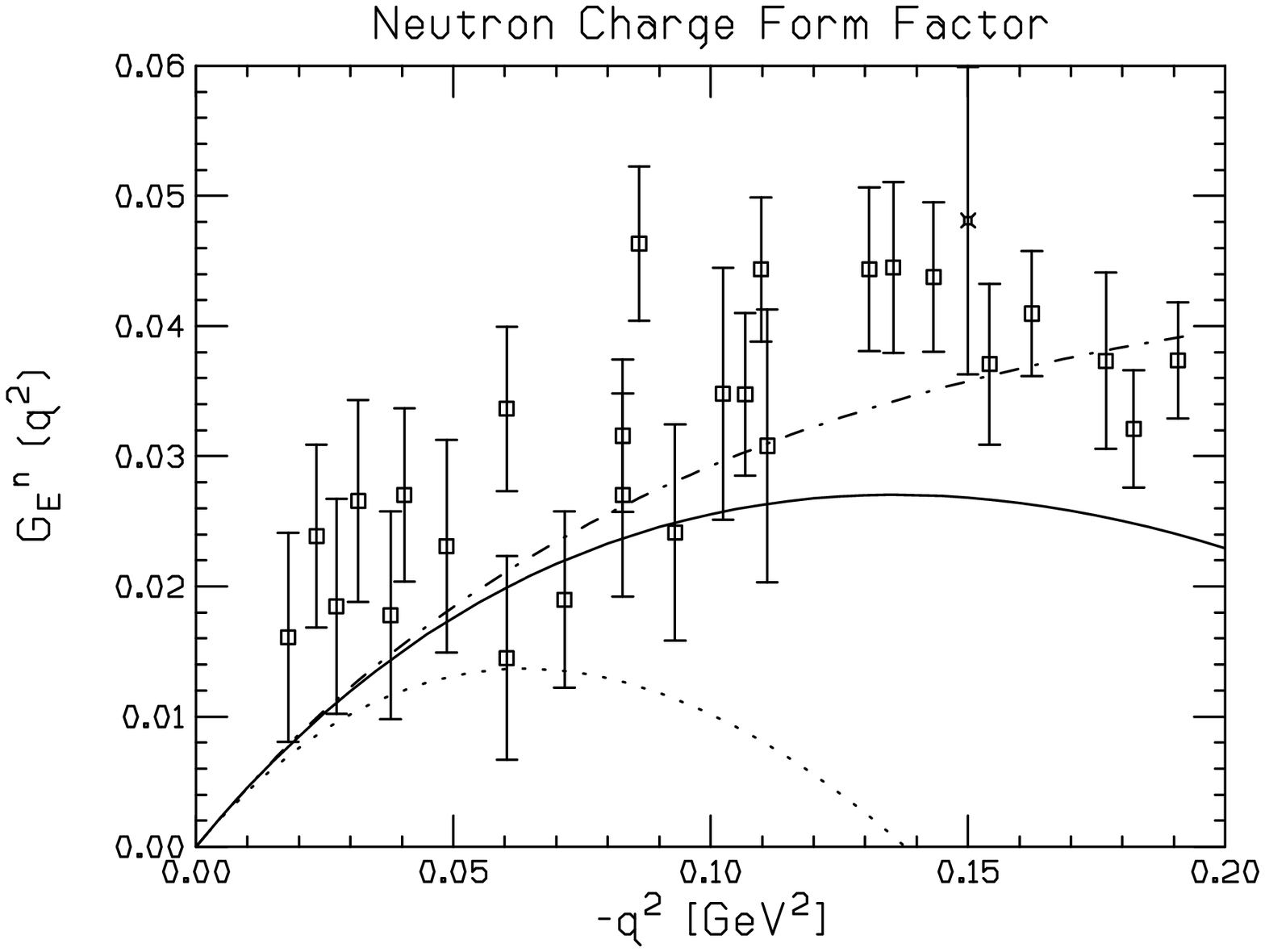,width=7.5cm}
\caption{Electric neutron form factor calculated within SU(3) (solid
  line) and SU(2) (dotted line). The dot--dashed
line is the result of the dispersive analysis, cf. fig.\ref{fig:Gen}.}
\label{fig:neusu3}
\end{minipage}
\end{figure}
\noindent Consider first the hyperons. The electric ffs of the charged hyperons
are given in fig.5. The corresponding radii are (a more detailed discussion
also of the neutral particles and magnetic radii is given in~\cite{khm})
\begin{equation}
\langle r^2_{\Sigma^+}\rangle = 0.64\ldots0.66~{\rm fm}^2~,\,
\langle r^2_{\Sigma^-}\rangle = 0.77\ldots0.80~{\rm fm}^2~,\,
\langle r^2_{\Xi^-}\rangle = 0.61\ldots0.65~{\rm fm}^2~.
\end{equation}
The given uncertainty does not reflect the contribution from higher orders,
which should be calculated. The prediction for the $\Sigma^-$ is in fair
agreement with the recent measurements.
The result for the $\Sigma$ radii is at variance
with quenched lattice QCD calculations which give 0.56(5)~fm$^2$ and
0.72(6)~fm$^2$ for the negative 
and positive $\Sigma$, respectively~\cite{lein}. However, quenched lattice calculations
should be taken with a grain of salt (the true error due to the quenching is
only known for very few quantities, certainly not for the radii). In the CHPT approach, 
the difference of the radii is due to some short distance physics
encoded in the LEC $d_{102}^0$ and to the Foldy term. The loop 
contributions are almost equal, but the difference due to the
counterterm and the Foldy term 
for the $\Sigma$ hyperons is
\begin{equation}
\langle r^2_{\Sigma^+}\rangle - \langle r^2_{\Sigma^-}\rangle =
-\frac{8d_{102}}{(4\pi F_\phi)^2} + \frac{b_D}{m^2} = -0.10 \ldots -0.15~{\rm fm}^2~,
\end{equation}
depending on how one fixes the electric LEC $d_{102}$ and the magnetic LEC $b_D$.
Here, $F_\phi = 100\,$MeV is the average pseudoscalar decay constant.
A more detailed discussion of the parameter dependence is given
in~\cite{khm}. All the numbers given here are based on a third order calculation. 
Clearly, a fourth order calculation is
called for to further quantify these results.

\noindent As can be seen from fig.6, the chiral description of the neutron
charge ff is clearly improved in SU(3) (solid line) as compared to the two
flavor case (dotted line). Obviously, this sizeable kaon cloud effect will be
reduced at next order since the effect of recoil only starts to show up
at fourth order. The inclusion of such recoil effects is expected 
to improve already the SU(2)
calculation, leaving less room for the kaon cloud effects. It is also worth
pointing out that this effect from kaon loops is opposite to what one expects from
a $\phi$--coupling~\cite{gk} and thus some cancellations should take place.

\section{Strange vector form factors}
\label{sec:strange}

\begin{figure}[htb]
\begin{minipage}[ht]{77mm}
\hspace{0.2cm}
\psfig{figure=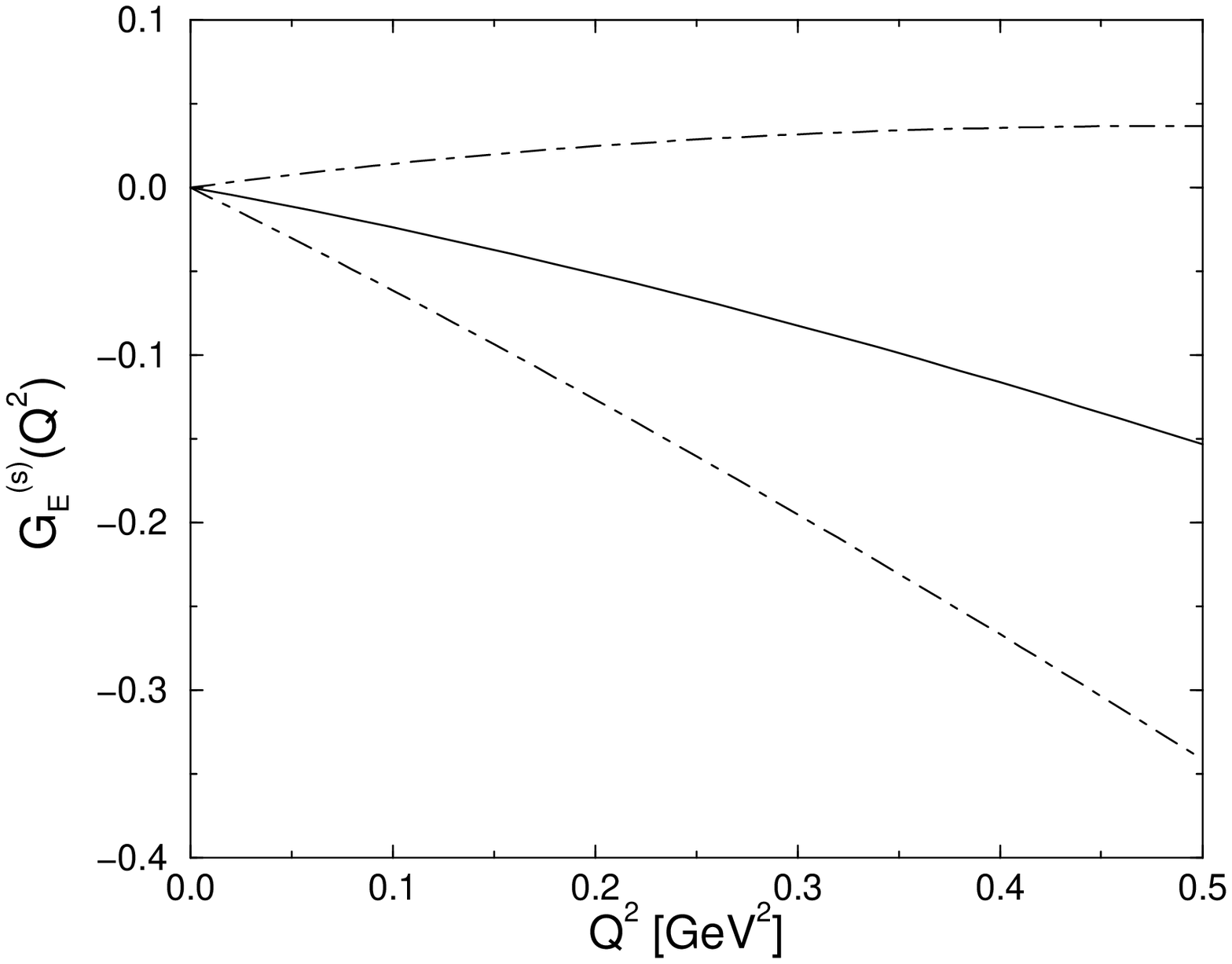,width=6.cm}
\caption{The strange electric form factor from chiral perturbation theory.}
\label{fig:estr}
\end{minipage}
\hspace{\fill}
\begin{minipage}[ht]{77mm}
\hspace{0.2cm}
\psfig{figure=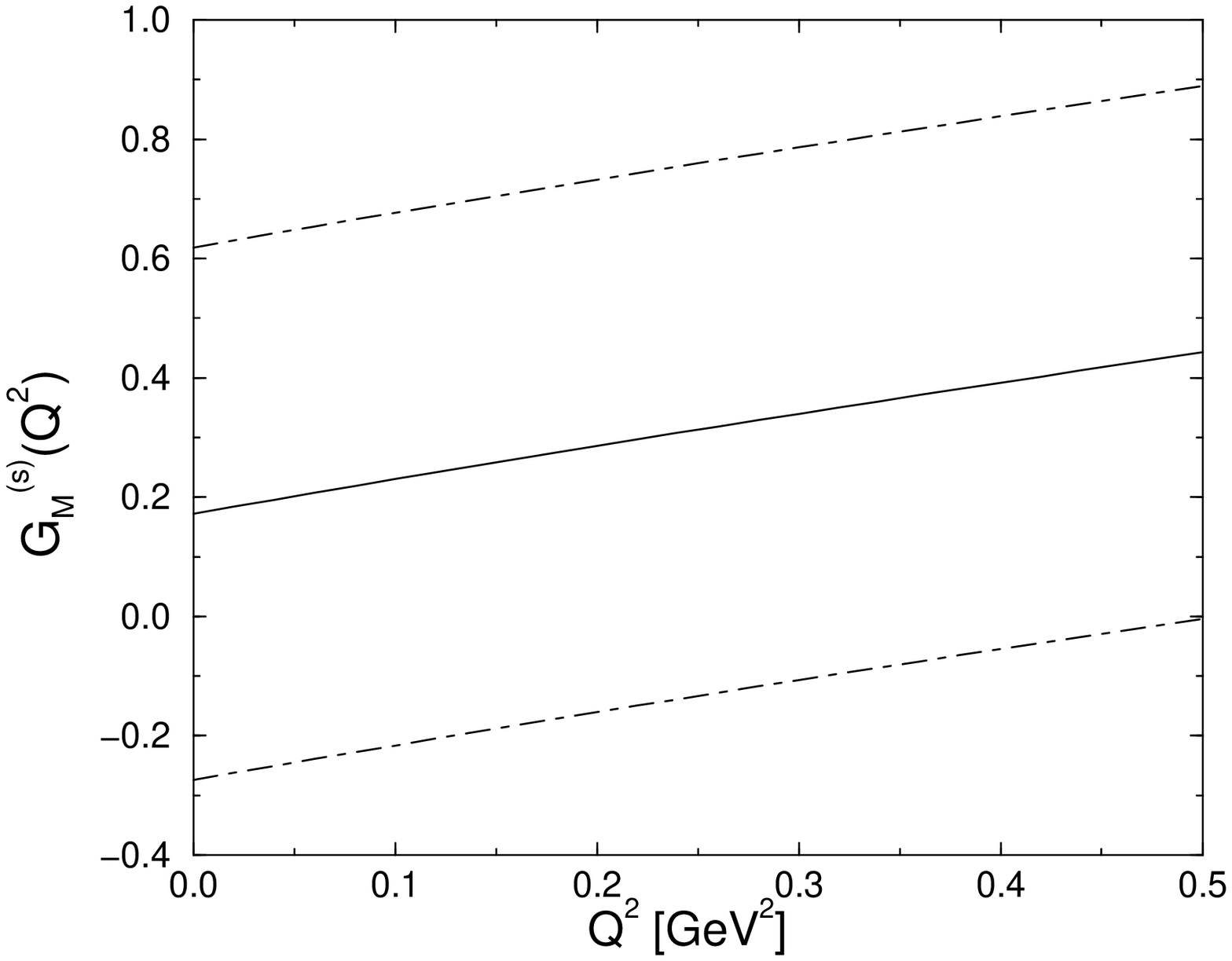,width=6.cm}
\caption{The strange magnetic form factor from chiral perturbation theory.}
\label{fig:mstr}
\end{minipage}
\end{figure}

\noindent 
Recently, the first results from parity--violating electron scattering
experiments, which allow to pin down the so--called strange form factors
of the nucleon, have become available. These strange ffs parametrize the
matrix elements of the strange vector current,
\begin{equation}\label{svc}
\langle N|\;\bar{s}\;\gamma_\mu\; s\;|N \rangle
= \langle N|\;\bar{q}\;\gamma_\mu\;
(\lambda^0/3-\lambda^8/\sqrt{3}) \; q\;| N \rangle~,
\end{equation}
with $q=(u,d,s)$ denoting the triplet of the light quark fields and
$\lambda^0 = I\; (\lambda^a)$ 
the unit (the $a=8$ Gell--Mann) SU(3) matrix. The singlet and octet currents
are parametrized in terms of  electric and magnetic ffs, which 
give the strange ffs via
\begin{equation}
G_{E/M}^{(s)} ( Q^{2}) = G_{E/M}^{(s)} (0) + \frac{1}{6}
\langle r^2_{E/M,s} \rangle \, Q^2 + {\cal O}(Q^4)~.
\end{equation}
The SAMPLE collaboration has reported the
first measurement of the strange magnetic moment of the proton~\cite{SAMPLE}. 
To be precise, they give the strange magnetic form factor in units of
nuclear magnetons at a small momentum transfer $Q_S^2=0.1~{\rm GeV}^2$,
$G_M^{(s)} (Q_S^2)= +0.23\pm 0.37 \pm 0.15\pm 0.19\,$.
The rather sizeable error bars document
the difficulty of such type of experiment.  The HAPPEX collaboration
has chosen a different kinematics which is more sensitive to the
strange electric form factor~\cite{HAPPEX}. Their measurement implies
$G_E^{(s)}(Q_H^2) + 0.39\;G_M^{(s)}
(Q_H^2) = 0.023 \pm 0.034 \pm 0.022\pm 0.026\,$,
at $Q_H^2=0.48~{\rm GeV}^2$. Of course, this momentum transfer might
be too high for the CHPT analysis at third order to hold, but in the absence
of data at lower $Q^2$ let us assume that we can still use the HAPPEX
result. This loophole should be kept in mind.
There have been many theoretical speculations about the size of the
strange form factors, some of them clearly in conflict with the
data.  These data have been analyzed in the framework of
chiral perturbation theory~\cite{hkm}, extending previous work~\cite{mus}.
 It was shown in~\cite{hms} that one
can make a parameter--free prediction for the momentum dependence of
the nucleons' strange magnetic (Sachs) form factor based on the chiral symmetry
of QCD solely. The value of the strange magnetic moment, which contains
an unknown low--energy constant ($b_0$), can be deduced from the SAMPLE experiment
using the momentum--dependence derived in~\cite{hms}. Furthermore, the SU(3)
analysis of the octet electromagnetic form factors performed in~\cite{khm}
allows one to pin down the octet component of the strange vector
current. Thus, to leading one--loop order, there is only
one new singlet counterterm ($d_{102}^0$), the strength of which can be determined from 
the value found by HAPPEX. This allows  to give a band for the
strange electric form factor and make a prediction for the MAMI A4
experiment, which intends to measure  
$G_E^{(s)}(Q_M^2) + 0.22\;G_M^{(s)}(Q_M^2)\;$
with a four-momentum transfer (squared) $Q^2_M=0.23~{\rm GeV}^2$
of approximately half the HAPPEX value.
Under the assumptions mentioned, one can determine the LECs
$b_0$ and $d_{102}^0$ with sizeable uncertainties reflecting the experimental
input. The central values are of natural size and the corresponding results
for the strange electric and magnetic ff are shown in 
figs.\ref{fig:estr},\ref{fig:mstr}
by the solid lines. The dashed lines reflect the theoretical uncertainty based
on a very conservative analysis. The corresponding strange radii and
the strange magnetic moment are~\cite{hms,hkm}
\begin{equation}
\langle r^2_{E,s} \rangle^{1/2} = (0.05 \pm 0.09) \, {\rm fm}^2~,\quad 
\langle r^2_{M,s} \rangle^{1/2} = -0.14  \, {\rm fm}^2~, 
\quad \mu_s = (0.18\pm 0.44)~{\rm n.m.}~,
\end{equation}
where the uncertainty in the strange radius stems mostly from the uncertainty
in the singlet LEC $d_{102}^0$, whereas the prediction for the magnetic radius
at this order is parameter--free. The uncertainty in $\mu_s$ is
completely given by the error of the SAMPLE analysis.
\noindent
A few more remarks on the strange electric ff are in order.
The radius is fairly small and {\it positive}, and even given the 
sizeable uncertainty, it is on the lower side 
of the predictions based on dispersive approaches including 
maximal OZI violation~\cite{bob,hmd2}. It is more compatible with models
that include $\pi \rho$~\cite{mmsvo} contributions in the isoscalar
spectral functions besides the vector
meson poles ($\omega, \phi, \ldots$)  or dispersive analysis of the
$\bar{K}K$~\cite{hm} continuum.
Note also that from the octet
current the strange electric radius inherits the  chiral singularity
$\sim \ln(M_K)$. It is also worth  pointing
out that the momentum dependence of the strange electric form factor
is rather different from the one of the neutron charge form factor,
which also vanishes at zero momentum transfer. We also note that using
the central values for the LECs, the prediction for the MAMI A4
experiment, which attempts to measure
$G_E^{(s)}(Q_M^2) + 0.22\;G_M^{(s)}(Q_M^2)$
at a four-momentum transfer (squared) of $Q^2_M=0.23~{\rm GeV}^2$, is
very small but afflicted with a large uncertainty. A more detailed discussion
is given in ref.\cite{hkm}. It is also important to stress that a
dispersive analysis of the $K\bar{K}$ continuum leads to  much smaller
values for the strange radii~\cite{hm2}. That approach is based on an
analytic continuation of the empirical $KN$ scattering amplitudes and
unitarity bounds are used. In principle, CHPT calculations and
dispersion relations can be mapped one--to--one as has been shown, see
e.g. ref.\cite{gm}. Both calculations need to be improved. On the CHPT
side, the next order has to be investigated in order  to check the convergence.
The dispersion relations need better data, since the analytic
continuation so far cannot be made stable without extra assumptions.
More direct data from Jefferson Lab and MAMI should help to clarify
the situation.

\section{Electromagnetic nucleon--delta transition form factors}

\noindent
Recently, a consistent scheme to include the $\Delta (1232)$ in a
chiral effective field theory was set up~\cite{hhk}. It is based on the
observation that the $N\Delta$ mass splitting $\delta = m_\Delta -m_N$
is only 300~MeV and that
the $\Delta$ is coupled strongly to the $N\pi\gamma$ system.
Treating the mass splitting as an additional small parameter, one then
expands in external momenta, pion mass insertions {\it and} $\delta$ (all
divided by the hadronic scale of 1~GeV). One
collectively denotes these small parameters as $\epsilon$. This
so--called small scale expansion  is a phenomenological extension
of CHPT. In ref.\cite{george}, the isovector $N\Delta$--transition was
calculated to third order in $\epsilon$. The transition matrix element
is parametrized in terms of three form factors $G_{1,2,3} (Q^2)$.
To that order, one has only two non--vanishing and finite loop
diagrams and all counterterms are momentum--independent. That means
that the $Q^2$--dependence of the transition ffs is predicted in a
parameter--free way and thus is a good testing ground for chiral dynamics.
Since the intermediate $\pi N$ state can go on mass--shell, these
ffs are complex, even at $Q^2=0$. That is not accounted for in most
models. These ffs can be mapped uniquely onto the multipole
amplitudes $M_1 (Q^2)$, $E_2 (Q^2)$ and $C_2 (Q^2)$. Consequently,
the $Q^2$--dependence of the EMR $E_2/M_1$ and of the CMR $C_2/M_1$
can be predicted. In figs.9,10 the momentum dependence of the EMR and
CMR is shown. The various lines refer to the multipole analysis of the
Mainz, RPI and VPI groups, which are used to pin down the LECs at
$Q^2=0$. A more detailed discussion of these topics can be found in 
ref.\cite{george}. It will be important to confront the recent
measurements from ELSA and BATES with these predictions.
\begin{figure}[htb]
\begin{minipage}[ht]{77mm}
\hspace{0.2cm}
\psfig{figure=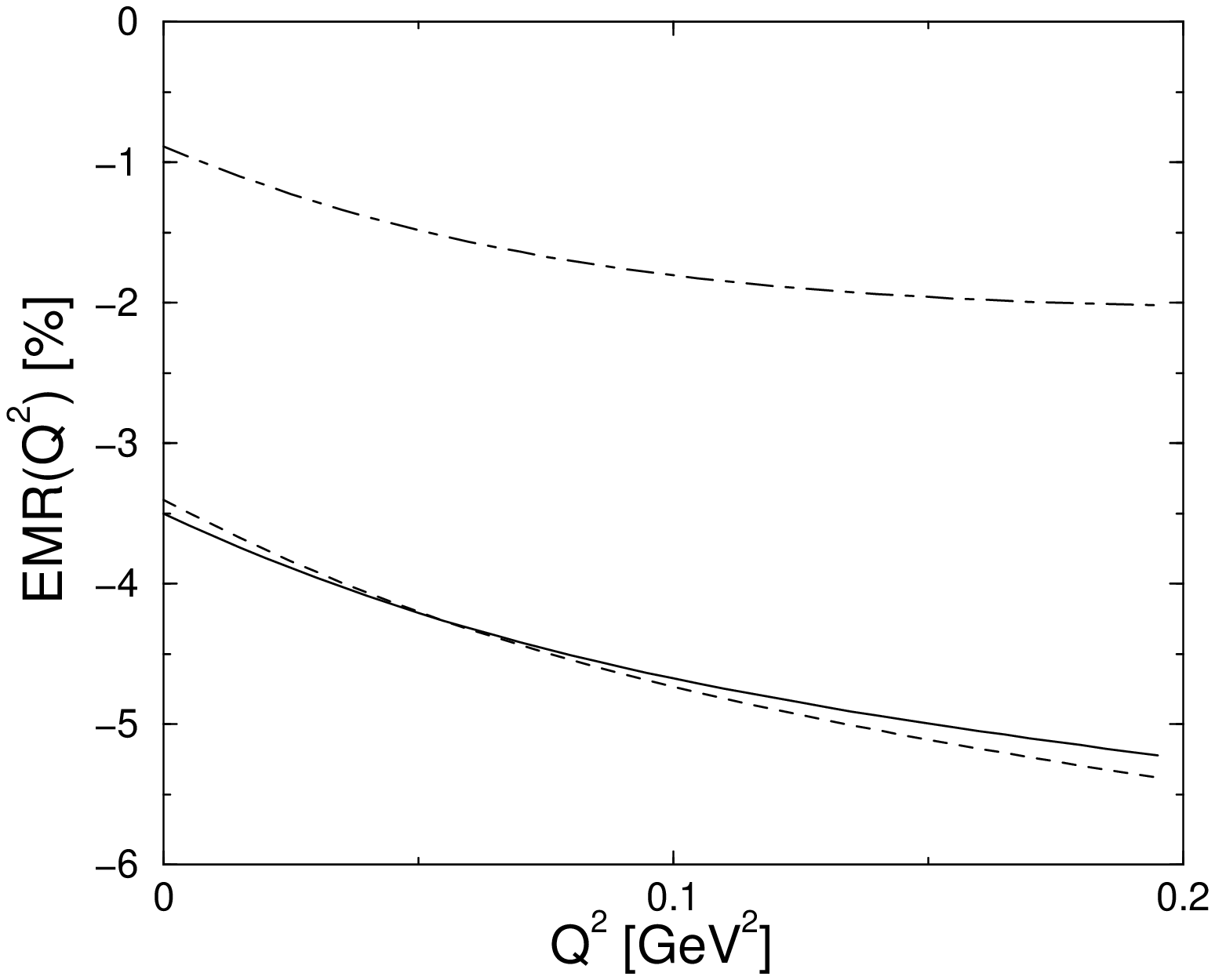,width=6.cm}
\vspace{-0.2cm}
\caption{The ratio $E_2/M_1$ versus $Q^2$. The solid, dashed and
  dot--dashed line refer to input from the Mainz, RPI and VPI
  multipole analysis, in order.}
\label{fig:emr}
\end{minipage}
\hspace{\fill}
\begin{minipage}[ht]{77mm}
\hspace{0.2cm}
\psfig{figure=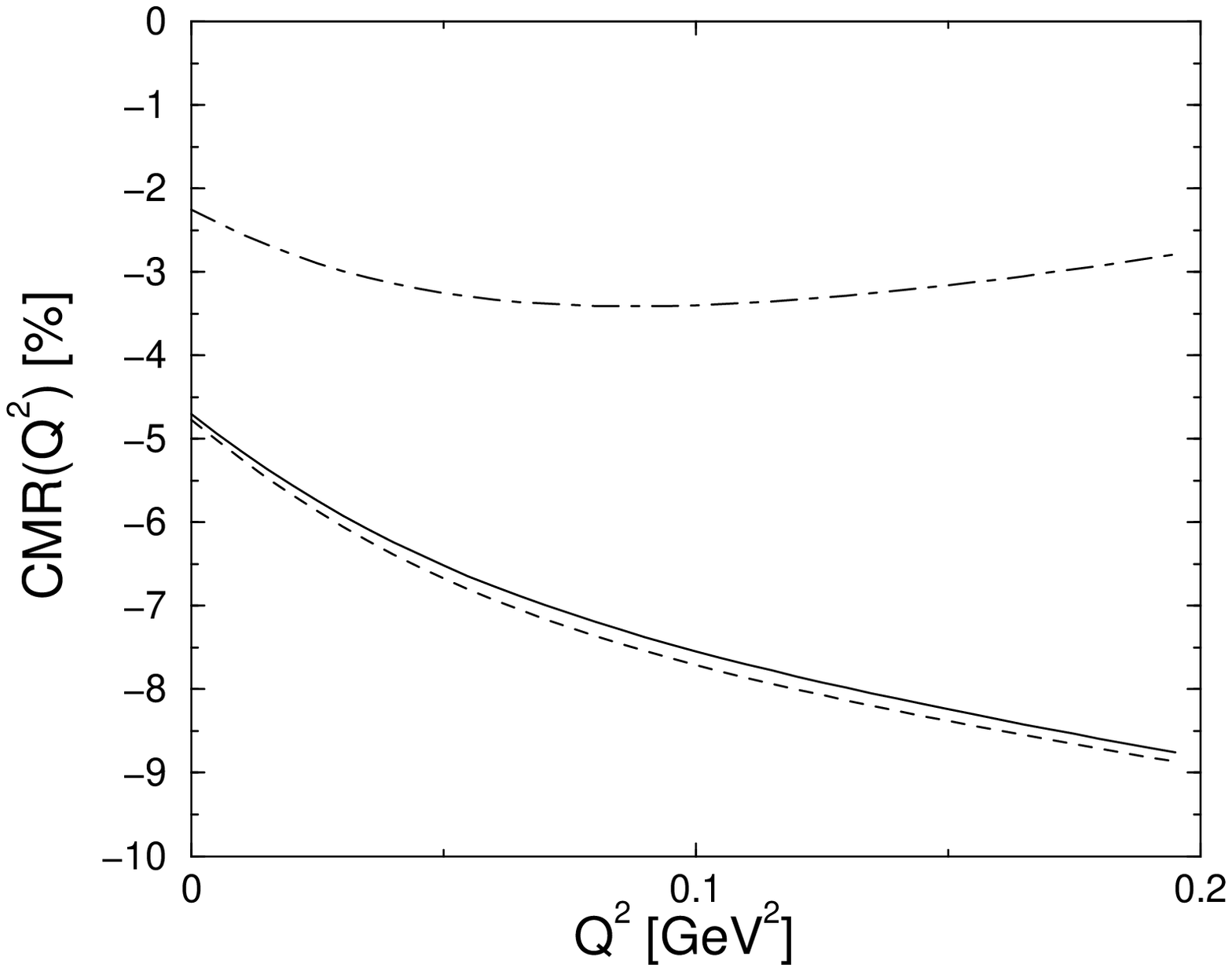,width=6.cm}
\vspace{-0.2cm}
\caption{The ratio $C_2/M_1$ versus $Q^2$. The solid, dashed and
  dot--dashed line refer to input from the Mainz, RPI and VPI
  multipole analysis, in order.}
\label{fig:cmr}
\end{minipage}
\end{figure}
\noindent

\section{Challenges}

\noindent To my opinion, there are three major issues to be resolved:

\begin{itemize}
\vspace{-0.2cm}

\item[$\star$] The dispersive analysis relies on a set of precise
and consistent data, otherwise the spectral functions can not be
extracted without severe assumptions. Clearly, the new data on
the ratio $\mu_p G_E^p (Q^2) /G_M^p(Q^2)$ cannot be explained for the
type of spectral functions used so far, consisting of a set of vector meson poles,
the $2\pi$--continuum and pQCD constraints. I would strongly encourage
studies implementing  other continua or correlated
multi--meson intermediate states in a consistent manner. 
Such approaches have been proven to be
fruitful in the study of baryon--baryon interactions. 
\vspace{-0.2cm}

\item[$\star$] A fourth order one--loop calculation of the electromagnetic
form factors of the baryon octet is mandatory. Such type of calculation
has already helped to deepen our understanding of the octet magnetic
moments, see ref.\cite{ms}. In particular, investigations of 
recoil effects in the neutron
electric form factor and the pattern of the hyperon charge radii are of
interest. A recently proposed Lorentz--invariant formulation
of baryon CHPT might prove to be a good tool~\cite{bl}.
\vspace{-0.2cm}

\item[$\star$] More theoretical work is needed to get a better handle
on the matrix elements of the strange vector current. With more data
particularly at low $Q^2$, a fourth order CHPT calculation can be attempted and
the effects of the spin-3/2 decuplet should be investigated~\cite{pm}.
Furthermore, the apparent discrepancy between the chiral prediction and the
one based on dispersion relations for the strange magnetic radius needs
to be resolved.

\end{itemize}

Finally, it is important to stress that all these problems are
intertwined. For example, the extraction of the strange ffs from
parity--violation experiments can only be done precisely if the data
are accurate enough but also the non--strange ffs are known precisely,
since in most parity--violation experiments the latter are used as
amplification factors.

\section*{Acknowledgements}

It is a pleasure to thank  my collaborators V\'eronique Bernard,
Harold Fearing,
Hans--Werner Hammer, Thomas Hemmert, Bastian Kubis and Sven Steininger.
Useful comments and communications from George Gellas,
Murray Moinester, Michael Ostrick
and Bob Wiringa are also acknowledged.

\end{document}